\documentclass[twocolumn,showpacs,preprintnumbers,amsmath,amssymb,superscriptaddress]{revtex4-1}
\usepackage{graphicx}% Include figure files
\usepackage{dcolumn}% Align table columns on decimal point
\usepackage{bm}% bold math

\begin{document}

\title{Thermoelectric internal current loops inside inhomogeneous systems}

\author{Y. Apertet}\email{yann.apertet@u-psud.fr}
\affiliation{Institut d'Electronique Fondamentale, Universit\'e Paris-Sud, CNRS, UMR 8622, F-91405 Orsay, France}
\author{H. Ouerdane}
\affiliation{CNRT Mat\'eriaux UMS CNRS 3318, 6 Boulevard Mar\'echal Juin, 14050 Caen Cedex, France}
\author{C. Goupil}
\affiliation{Laboratoire CRISMAT, UMR 6508 CNRS, ENSICAEN et Universit\'e de Caen Basse Normandie, 6 Boulevard Mar\'echal Juin, 14050 Caen, France}
\author{Ph. Lecoeur}
\affiliation{Institut d'Electronique Fondamentale, Universit\'e Paris Sud CNRS, 91405 Orsay, France, CNRS, UMR 8622, 91405 Orsay, France}

\date{\today}

\begin{abstract}
Considering a system composed of two different thermoelectric modules electrically and thermally connected in parallel, we demonstrate that the inhomogeneities of the thermoelectric properties of the materials may cause the appearance of an electrical current, which develops inside the system. We show that this current increases the effective thermal conductance of the whole system. We also discuss the significance of a recent finding concerning a reported new electrothermal effect in inhomogeneous bipolar semiconductors, in light of our results.
\end{abstract}


\maketitle

Thermoelectric power generation is a promising way to achieve efficient waste energy harvesting. To ensure a high heat-to-electrical power conversion efficiency, the thermal conductances of the materials used for thermoelectric modules (TEM) have, in principle, to be as low as possible\cite{Shakouri2011}. Fu \emph{et al} \cite{Fu2011} recently reported on an electrothermal process that can modify the effective thermopower of semiconductor devices. In particular, they claimed that the joint application of a temperature gradient and an electric field (perpendicular to each other) to a bipolar semiconductor structure induces steady current vortices (even in open circuit configuration) which in turn yield Joule heating whose effect is to lower thermal conductivity. It is thus worthwhile to check whether this effect may be used to improve the so-called figure of merit $ZT$ of bipolar semiconductor structures to a significant degree.

The theoretical prediction of Fu \emph{et al} \cite{Fu2011} that internal current vortices formed at a pn junction provide a way to reduce thermal conductivity in practical devices deserves closer inspection. In this Brief Report, using a macroscopic description of a two-leg TEM, we demonstrate that an internal current also gives rise to \emph{advective thermal transport} which, unfortunately for practical applications, largely compensates the effect proposed by Fu \emph{et al} \cite{Fu2011} and hence effectively lowers $ZT$. Studying the simple case of two thermoelectric modules connected in parallel both thermally and electrically, we suggest that internal currents caused by a temperature gradient are not directly linked to the transverse electrical field as supposed in Ref.~\cite{Fu2011} but rather caused by thermoelectric inhomogeneities inside the materials.

\begin{figure}
	\centering
		\includegraphics[width=0.45\textwidth]{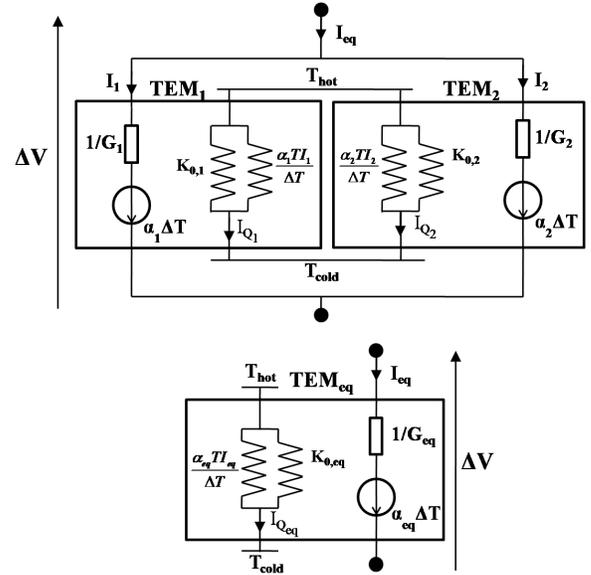}
	\caption{Coupled thermoelectric modules (top) and equivalent module (bottom).}
	\label{fig:figure1}
\end{figure}

To gain insight into the main features of internal current loops, let us consider two thermoelectric modules TEM$_1$ and TEM$_2$, and the equivalent module, denoted TEM$_{\rm eq}$, resulting from their association in parallel, both electrically and thermally as shown in Fig.~\ref{fig:figure1}. Each of them is characterized by its isothermal electrical conductance $G_{i}$, its thermal conductance under open electrical circuit condition $K_{0,i}$ and its Seebeck coefficient $\alpha_{i}$, where $i$ can be $1$,$2$ or ${\rm eq}$ as appropriate. All these coefficients are supposed constant. The whole system is subjected to a temperature difference $\Delta T=T_{\rm hot}-T_{\rm cold}$, and its average temperature is $T$.

Using linear response theory we can express each electrical current $I_i$ and thermal flux $I_{Q_{_{i}}}$ as functions of generalized forces related to the temperature difference $\Delta T$ and voltage difference $\Delta V$ to which the TEM is subjected. The potential differences $\Delta V$ and $\Delta T$ are the same for TEM$_1$ and TEM$_2$ and hence for TEM$_{\rm eq}$ by construction. The relation between the fluxes and the forces is given by \cite{Callen1948}:

\begin{equation}\label{frcflx}
\left(
\begin{array}{c}
I_i\\
I_{Q_{_{i}}}\\
\end{array}
\right)
=
G_i
\left(
\begin{array}{cc}
1~ & ~\alpha_i\\
\alpha_i T~ & ~\alpha_i^2 T + K_{0,i}/G_i\\
\end{array}
\right)
\left(
\begin{array}{c}
\Delta V\\
\Delta T\\
\end{array}
\right),
\end{equation}

\noindent from which we obtain a quite simple expression of the thermal flux: 

\begin{equation}
\label{eq:IQI}
I_{Q_{_{i}}}=\alpha_i T I_i+K_{0,i}\Delta T,
\end{equation}

\noindent This equation shows that two distinct processes contribute to the thermal transport: one is linked to thermal conduction by both phonons and electrons when there is no current flowing inside the structure (term in $K_0\Delta T$), the other to electrical current flow (term in $\alpha T I$). Since this second contribution is associated with a macroscopic displacement of electrons it can be stated to be thermal transport by {\it electronic advection}; and the heat quantity transported by each electron \cite{Pottier2007} is given by $|\alpha| T e$, $e$ being the elementary electric charge. This notion of {\it electronic advection} is central to explain the increase of thermal conductance when an internal current develops inside the structure. We stress that the additional term should not be confused with the electronic part of $K_0$, which is used, for example, in the Wiedemann-Franz law. On Fig.~\ref{fig:figure1}, this is shown with the added thermal conductance parameterized by the electrical current.

Besides constitutive laws for each module given by Eq.~(\ref{frcflx}), there are additionnal relations linked to the parallel configuration that must be accounted for. First, we ensure electrical current conservation:

\begin{equation}
\label{eq:conservationI}
I_{\rm eq}=I_1+I_2
\end{equation}

\noindent Next, we consider that the mean thermal flux flowing through TEM$_{\rm eq}$ is the sum of the two mean thermal fluxes flowing through TEM$_1$ and TEM$_2$; hence

\begin{equation}
\label{eq:conservationIQ}
I_{Q_{_{\rm eq}}}=I_{Q_{_{1}}}+I_{Q_{_{2}}}
\end{equation}

\noindent Now using Eqs.~(\ref{frcflx}), (\ref{eq:conservationI}) and (\ref{eq:conservationIQ}) we are free choose the thermal and electrical configurations which permits an easy derivation of the equivalent parameters of the system as a whole. Under isothermal condition, $\Delta T=0$, and Eq.~(\ref{eq:conservationI}) leads to:

\begin{equation}
\label{eq:conductanceeff}
G_{\rm eq}=G_1 +G_2,
\end{equation}

Under closed circuit condition, $\Delta V=0$, and Eq.~(\ref{eq:conservationI}) leads to $G_{\rm eq}\alpha_{\rm eq}=G_{1}\alpha_{1}+G_{2}\alpha_{2}$, so that the equivalent thermopower $\alpha_{\rm eq}$ is defined as the weighted average of the two Seebeck coefficients $\alpha_{1}$ and $\alpha_{2}$:

\begin{equation}
\label{eq:alphaeff}
\alpha_{\rm eq}=\frac{G_1 \alpha_1 + G_2 \alpha_2}{G_1 + G_2} ,
\end{equation}

\noindent This equation is the same as the one given by Hicks and Dresselhaus (Eq.~(2) of Ref.~\cite{Hicks1993}) for a semiconductor with two conduction bands. The correspondence between both is explained by the fact that each conduction band can be associated with a medium where electrons flow parallelly. From this result we see that the effective Seebeck effect cannot exceed the larger one of the two materials.

The equivalent thermal conductance is now determined under open circuit condition, $I_{\rm eq}=0$. For nonzero values of $I_1$ and $I_2$, this condition is satisfied for $I_1=-I_2$; since the conservation of the thermal flux (\ref{eq:conservationIQ}) remains valid, we obtain:

\begin{equation}
\label{eq:condthermeff0}
K_{0,{\rm eq}}=\frac{(\alpha_1 -\alpha_2) T I_1}{\Delta T}+K_{0,1}+K_{0,2},
\end{equation}

\noindent using Eq.~(\ref{eq:IQI}). To proceed, we determine the intensity $I_1$ as follows. Under open ciruit condition $I_{\rm eq}=0$ so that the voltage reads $\Delta V=-\alpha_{\rm eq}\Delta T$, which we include in the expression of $I_1$ given by \eqref{frcflx} to obtain:

\begin{equation}
\label{eq:I1}
I_1=\frac{G_1G_2}{G_1+G_2}(\alpha_1 - \alpha_2)\Delta T
\end{equation}

\noindent  This equation shows that there is a non-zero electrical current flowing as long as the two Seebeck coefficients are different. Now, the substitution of the obtained expression of $I_1$ into Eq.~(\ref{eq:condthermeff0}), yields the equivalent thermal conductance at zero current:

\begin{equation}
\label{eq:condthermeff}
K_{0,{\rm eq}}=K_{0,1}+K_{0,2}+\frac{G_1G_2}{G_1+G_2}(\alpha_1 - \alpha_2)^2T
\end{equation}

\noindent The above formula exibits an additional term next to the sum of the thermal conductance of each module. This term is related to the internal current that develops inside the structure when sumitted to a temperature gradient. This current is proportional to the difference between the Seebeck coefficients of each leg. The total transported heat by this current is proportionnal to  this difference too, so the increase in thermal conductance is proportionnal to $(\alpha_1-\alpha_2)^2$.

As an internal current is generated, energy dissipation is caused by Joule effect. The total dissipated power is given by the sum of the power dissipated in each part of the whole system: $P_{\rm Joule} = I_{\rm int}^2(1/G_1+1/G_2)$, where $I_{\rm int}~(\equiv I_1)$ is given by Eq.~(\ref{eq:I1}); an explicit expression is,

\begin{equation}\label{eq:Pjoule}
P_{\rm Joule}=\frac{G_1G_2}{G_1+G_2}(\alpha_1 - \alpha_2)^2\Delta T^2.
\end{equation}

\noindent The dependence on the square of the difference of the Seebeck coefficients of the modules shows that the more they are dissimilar in terms of thermopower the more Joule dissipation is important. Here, there is no need to assume, as one would for the classical generator with two legs, that each module has a different doping type. We also note that the internal current is stronger for materials with high electrical conductance, hence metals are more sensitive to small inhomogeneities in the Seebeck coefficient. This explains why this effect is exploited in non-destructive testing to probe metallic inclusions in a host metal \cite{Nayfeh2002,Kleber2005,Carreon2003}.

Let us remark that for inclusions at the nanoscale the previous statement about the increase of the thermal conductance no longer stands since these inclusions have a strong impact on thermal conduction by phonons. For example Kim \emph{et al} \cite{Kim2006} demonstrated that nanoinclusions are efficient to lower the thermal conductivity in InGaAs.

We now turn to the analysis of the paper of Fu \emph{et al} \cite{Fu2011} using our Eq.~(\ref{eq:condthermeff}), which allows to explain the behavior of a pn junction submitted to a transverse thermal gradient under open circuit condition. A typical Pisarenko plot \cite{Ioffe1957}, i.e. the Seebeck coefficient plotted against the carrier concentration, shows that the Seebeck coefficient is higher for low carrier density. In a pn junction a depleted zone forms and develops on each side of the interface over a few hundreds of nanometers depending on the doping concentration: in these two regions, the Seebeck coefficients increase locally and thus become greater than those in the quasi-neutral regions. Since the carrier concentration inhomogeneity is transverse to the applied temperature gradient, this situation is similar to the simple one studied in the present paper and we can expect internal current vortices to develop as described in \cite{Fu2011}. We believe that the electric field present in the space charge zone is a consequence of the carrier depletion and is not a cause \emph{per se} of the internal current generation: in a structure where one can tune the Seebeck coefficient without changing the carrier density we expect to find the same behavior. However, for the particular case of a pn junction the Seebeck inhomogeneity and the transverse electric field are closely connected through the carrier depletion zone; therefore linking, in this case, the internal current either to thermoelectric inhomogeneties or to a transverse electric field essentially amounts to express two viewpoints on the same phenomenon. Besides, the appearence of two vortices, one on each side of the junction, is due to the potentiel barrier arising at the interfaces: each type of carrier is confined in its own side so that these two separate systems have no influence on each other except when one of the sides becomes thinner than the depleted zone.

We state further that a decrease of the thermal conductance is not possible in such circumstance: Fu and co-workers \cite{Fu2011} overlooked the advective part of the thermal flux. According to these authors, the reduction is due to the fact that half of the energy dissipated by the Joule effect is actually going back to the hot side. This heat quantity sould be compared to the one transported by \emph{electronic heat advection}: the ratio between these two quantities scales as $\Delta T/2T$. In the framework of linear theory, the temperature difference $\Delta T$ is assumed to be small, so the heat flowing back to the hot reservoir is small compared to the \emph{advection} part: internal currents increase the thermal conductance; they do not decrease it. Generation of internal currents in spatially inhomogeneous systems such as multilayer structures were discussed twenty years ago by Saleh \emph{et al} \cite{Saleh1991}; our analysis is consistent with their conclusions.

The specific case of TEMs with both electrical and thermal parallel configurations satisfies Bergman's theorem, which states that the figure of merit $ZT$ of a composite material cannot be greater than the larger $ZT$ of the consituents \cite{Bergman1991}. We have shown that the Seebeck coefficient is lower than the larger one of the two, Eq.~(\ref{eq:alphaeff}). The effective electrical conductance of two conductances in parallel is always greater than the larger one, but, since $ZT\propto G/K_0$ for a give temperature $T$, this increase is counterbalanced by the fact the thermal conductances behave similarly when no internal current develops. Moreover, considering a TEM designed with components that have different Seebeck coefficients, the increase in equivalent thermal conductance is even stronger. So, as expressed by Bergman and Levy \cite{Bergman1991}, the figure of merit of the TEM$_{\rm eq}$ can only be lower than the highest $ZT$ of the more efficient TEM. Indeed, the addition of an internal current can only lower the figure of merit, since it adds a dissipative process to the system. A full generalization of this result to composite materials necessitates the derivation of an expression of the effective Seebeck coefficient for a configuration where TEMs are both thermally and electrically in series. Since a composite system may be viewed as a network of TEMs a general formulation can, in principle, be obtained; however we anticipate that such derivation, which is beyond the scope of the present Brief Report, can be quite tricky. The simple example given in the present paper suffices to gain insight into the problem of internal currents and their properties.

To end this Brief Report, we make two additional remarks: 

Assuming that the properties of the thermoelectric modules are temperature-dependent, it would be interesting to check if it is possible to create instabilities inside the module analogous to the Rayleigh-B\'enard convection phenomenon which occurs in conventional fluids subjected to a thermal gradient.

The present work and the cited ones on current loops bring us back to the original interpretation of the thermoelectric effect made by Seebeck who thought that the temperature difference led to magnetism whereas it was the internal current of the structure that created the magnetic field, which deviated the compass needle~\cite{Seebeck}.\\

\paragraph*{Acknowledgments}
This work is part of the CERES 2 and ISIS projects funded by the Agence Nationale de la Recherche. Y. A. acknowledges financial support from the Minist\`ere de l'Enseignement Sup\'erieur et de la Recherche.

\end{document}